\documentclass[aps,twocolumn,floatfix,showpacs,superscriptaddress]{revtex4}
\usepackage{graphicx}
\usepackage{amsmath}
\usepackage{amssymb}
\usepackage{bm}

\usepackage{color}

\begin{document}

\title{Coulomb-assisted braiding of Majorana fermions in a Josephson junction array}
\author{B. van Heck}
\affiliation{Instituut-Lorentz, Universiteit Leiden, P.O. Box 9506, 2300 RA Leiden, The Netherlands}
\author{A. R. Akhmerov}
\affiliation{Instituut-Lorentz, Universiteit Leiden, P.O. Box 9506, 2300 RA Leiden, The Netherlands}
\author{F. Hassler}
\affiliation{Institute for Quantum Information, RWTH Aachen University, D-52056 Aachen, Germany}
\author{M. Burrello}
\affiliation{Instituut-Lorentz, Universiteit Leiden, P.O. Box 9506, 2300 RA Leiden, The Netherlands}
\author{C. W. J. Beenakker}
\affiliation{Instituut-Lorentz, Universiteit Leiden, P.O. Box 9506, 2300 RA Leiden, The Netherlands}

\date{November, 2011}

\begin{abstract}
We show how to exchange (braid) Majorana fermions in a network of superconducting nanowires by control over Coulomb interactions rather than tunneling. Even though Majorana fermions are charge-neutral quasiparticles (equal to their own antiparticle), they have an effective long-range interaction through the even-odd electron number  dependence of the superconducting ground state. The flux through a split Josephson junction controls this interaction via the ratio of Josephson and charging energies, with exponential sensitivity. By switching the interaction on and off in neighboring segments of a Josephson junction array, the non-Abelian braiding statistics can be realized without the need to control tunnel couplings by gate electrodes. This is a solution to the problem how to operate on topological qubits when gate voltages are screened by the superconductor.
\end{abstract}

\maketitle

\section{Introduction}
\label{intro}

Non-Abelian anyons have a topological charge that provides a nonlocal encoding of quantum information \cite{Nay08}. In superconducting implementations \cite{Lut10,Ore10} the topological charge equals the electrical charge modulo $2e$, shared nonlocally by a pair of midgap states called Majorana fermions \cite{Kit01}. This mundane identification of topological and electrical charge by no means diminishes the relevance for quantum computation. To the contrary, it provides a powerful way to manipulate the topological charge through the well-established sub-$e$ charge sensitivity of superconducting electronics \cite{Ave92,Dev04}.

Following this line of thought, three of us recently proposed a hybrid device called a \textit{top-transmon}, which combines the adjustable charge sensitivity of a superconducting charge qubit (the \textit{transmon} \cite{Sch08}) to read out and rotate a topological (\textit{top}) qubit \cite{Has11}. A universal quantum computer with highly favorable error threshold can be constructed \cite{Bra05} if these operations are supplemented by the pairwise exchange (braiding) of Majorana fermions, which is a non-Abelian operation on the degenerate ground state \cite{Rea00,Iva01}.

Here we show how Majorana fermions can be braided by means of charge-sensitive superconducting electronics. (Braiding was not implemented in Ref.\ \cite{Has11} nor in other studies of hybrid topological/nontopological superconducting qubits \cite{Has10,Sau10,Fle11,Jia11,Bon11}.) We exploit the fact that the charge-sensitivity can be switched on and off \textit{with exponential accuracy} by varying the magnetic flux through a split Josephson junction \cite{Sch08}. This provides a macroscopic handle on the Coulomb interaction of pairs of Majorana fermions, which makes it possible to transport and exchange them in a Josephson junction array.

We compare and contrast our approach with that of Sau, Clarke, and Tewari, who showed (building on the work of Alicea et al.\ \cite{Ali11}) how non-Abelian braiding statistics could be generated by switching on and off the tunnel coupling of adjacent pairs of Majorana fermions \cite{Sau11}. The tunnel coupling is controlled by a gate voltage, while we rely on Coulomb interaction controlled by a magnetic flux. This becomes an essential difference when electric fields are screened too strongly by the superconductor to be effective. (For an alternative non-electrical approach to braiding, see Ref.\ \cite{Rom11}.)

The basic procedure can be explained quite simply, see Sec.\ \ref{braiding}, after the mechanism of the Coulomb coupling is presented in Sec.\ \ref{MCH}. We make use of two more involved pieces of theoretical analysis, one is the derivation of the low-energy Hamiltonian of the Coulomb coupled Majorana fermions (using results from Refs.\ \cite{Fu10,Hec11}), and the other is the calculation of the non-Abelian Berry phase \cite{Wil84} of the exchange operation. To streamline the paper the details of these two calculations are given in Appendices.

\section{Majorana-Coulomb Hamiltonian}
\label{MCH}

\subsection{Single island}
\label{single_island}

\begin{figure}[tb] 
\centerline{\includegraphics[width=0.6\linewidth]{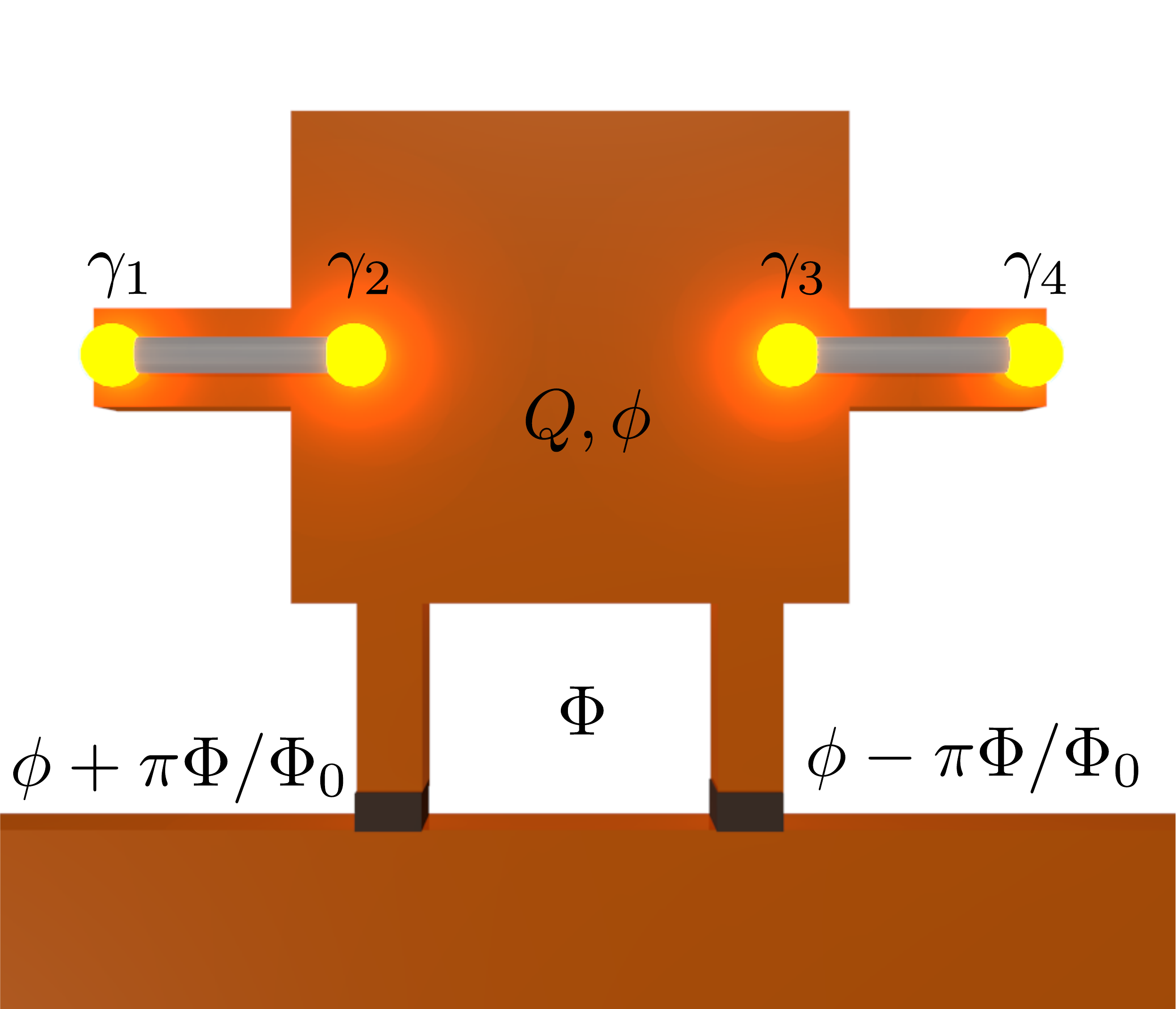}}
\caption{\label{fig_box}
Cooper pair box, consisting of a superconducting island (brown) connected to a bulk superconductor by a split Josephson junction (black, with the gauge-variant phase differences indicated). The island contains Majorana fermions (yellow) at the end points of a nanowire (grey). These are coupled by the Coulomb charging energy, tunable via the flux $\Phi$ through the Josephson junction.
}
\end{figure}

The basic building block of the Josephson junction array is the Cooper pair box \cite{Mak01}, see Fig.\ \ref{fig_box}, consisting of a superconducting island (capacitance $C$) connected to a bulk (grounded) superconductor by a split Josephson junction enclosing a magnetic flux $\Phi$. The Josephson energy $E_{J}$ is a periodic function of $\Phi$ with period $\Phi_{0}=h/2e$. If the two arms of the split junction are balanced, each with the same coupling energy $E_{0}$, the Josephson energy
\begin{equation}
E_{J}=2E_{0}\cos(\pi\Phi/\Phi_{0})\label{EJE0}
\end{equation}
varies between $0$ and $2E_{0}>0$ as a function of $|\Phi|<\Phi_{0}/2$.

When the island contains no Majorana fermions, its Hamiltonian has the usual form \cite{Tin96}
\begin{equation}
H=\frac{1}{2C}(Q+q_{\rm ind})^{2}-E_{J}\cos\phi,\label{Hsingle}
\end{equation}
in terms of the canonically conjugate phase $\phi$ and charge $Q=-2ei\,d/d\phi$ of the island. The offset $q_{\rm ind}$ accounts for charges on nearby gate electrodes. 
We have chosen a gauge such that the phase of the pair potential is zero on the bulk superconductor.

A segment of a semiconductor nanowire (typically InAs) on the superconducting island can have Majorana midgap states bound to the end points \cite{Lut10,Ore10}. For $N$ segments there can be $2N$ Majorana fermions on the island. They have identical creation and annihilation operators $\gamma_{n}=\gamma_{n}^{\dagger}$ satisfying
\begin{equation}
\gamma_{n}\gamma_{m}+\gamma_{m}\gamma_{n}=2\delta_{nm}.\label{gammacom} 
\end{equation}
The topological charge of the island equals the fermion parity
\begin{equation}
{\cal P}=i^{N}\prod_{n=1}^{2N}\gamma_{n}.\label{Pdef}
\end{equation}
The eigenvalues of ${\cal P}$ are $\pm 1$, depending on whether there is an even or an odd number of electrons on the island. 

The Majorana operators do not enter explicitly in $H$, but affect the spectrum through a constraint on the eigenstates \cite{Fu10},
\begin{equation}
\Psi(\phi+2\pi)=(-1)^{(1-{\cal P})/2}\Psi(\phi).\label{Psiphi}
\end{equation}
This ensures that the eigenvalues of $Q$ are even multiples of $e$ for ${\cal P}=1$ and odd multiples for ${\cal P}=-1$. Since ${\cal P}$ contains the product of all the Majorana operators on the island, the constraint \eqref{Psiphi} effectively couples distant Majorana fermions --- without requiring any overlap of wave functions.

We operate the Cooper pair box in the regime that the Josephson energy $E_{J}$ is large compared to the single-electron charging energy $E_{C}=e^{2}/2C$. The phase $\phi$ (modulo $2\pi$) then has small zero-point fluctuations around the value $\phi_{\rm min}=0$ which minimizes the energy of the Josephson junction, with occasional $2\pi$ quantum phase slips. 

In Appendix \ref{Cinteraction} we derive the effective low-energy Hamiltonian for $E_{J}\gg E_{C}$,
\begin{align}
&H_{\rm eff}=-E_{J}+\sqrt{2 E_CE_J}-U{\cal P},\label{Heff}\\
&U=16(E_{C}E_{J}^{3}/2\pi^{2})^{1/4}e^{-\sqrt{8E_{J}/E_{C}}}\cos(\pi q_{\rm ind}/e).\label{Udef}
\end{align}
The energy minimum $-2E_{0}$ at $\phi_{\rm min}$ is increased by $\sqrt{2E_CE_J}$ due to zero-point fluctuations of the phase. This offset does not contain the Majorana operators, so it can be ignored. The term $-U{\cal P}$ due to quantum phase slips depends on the Majorana operators through the fermion parity. This term acquires a dynamics for multiple coupled islands, because then the fermion parity of each individual island is no longer conserved.

\subsection{Multiple islands}
\label{multiple_islands}

\begin{figure}[tb] 
\centerline{\includegraphics[width=0.8\linewidth]{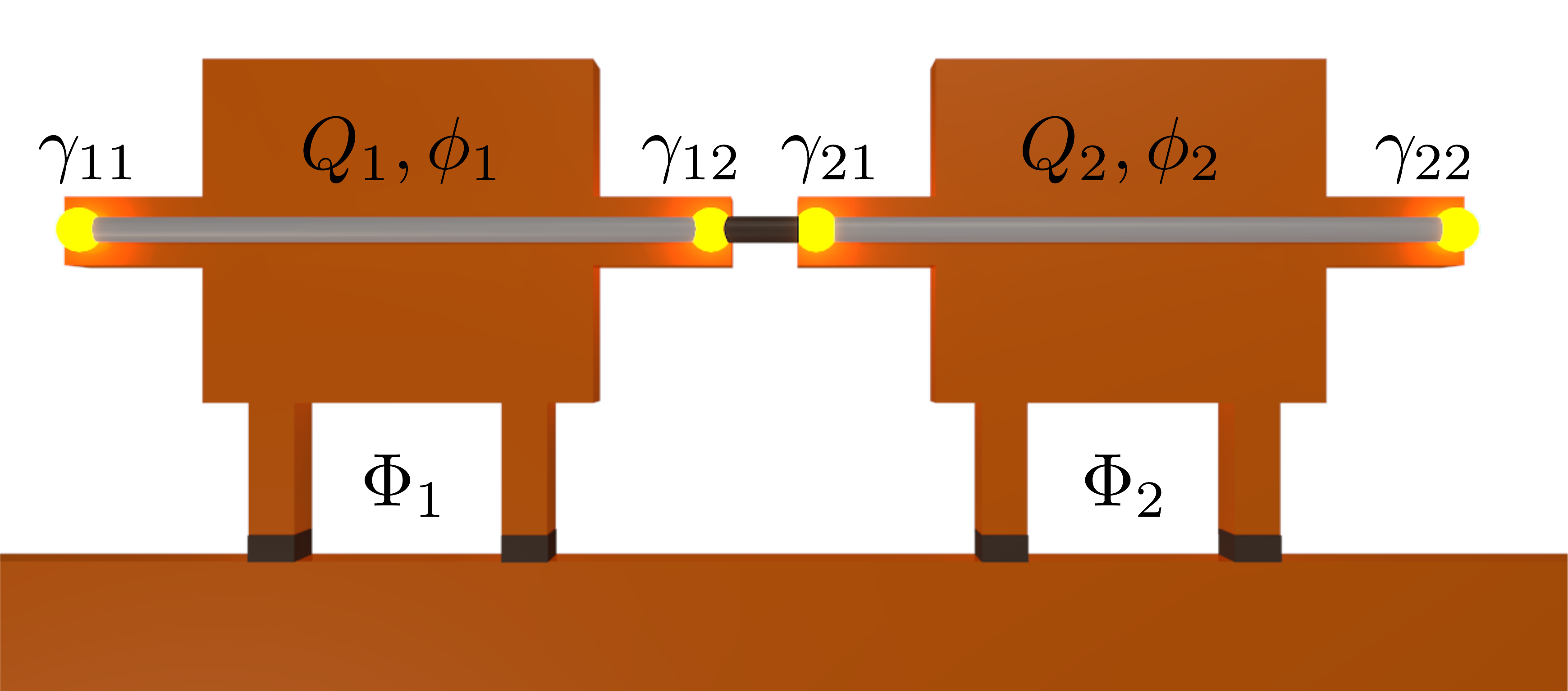}}
\caption{\label{fig_islands} 
Two Cooper pair boxes, each containing a pair of Majorana fermions. Single electrons can tunnel between the superconducting islands via the overlapping Majorana's $\gamma_{12}$ and $\gamma_{21}$. This tunnel coupling has a slow (cosine) dependence on the enclosed fluxes, while the Coulomb coupling between the Majorana's on the same island varies rapidly (exponentially).
}
\end{figure}

We generalize the description to multiple superconducting islands, labeled $k=1,2,\ldots$, each connected to a bulk superconductor by a split Josephson junction enclosing a flux $\Phi_{k}$. (See Fig.\ \ref{fig_islands}.) The Josephson junctions contribute an energy 
\begin{equation}
H_{J}=-\sum_{k}E_{J,k}\cos\phi_{k},\;\;E_{J,k}=2E_{0}\cos(\pi\Phi_{k}/\Phi_{0}).\label{HJdef}
\end{equation}
We assume that the charging energy is dominated by the self-capacitance $C$ of each island, so that it has the additive form
\begin{equation}
H_{C}=\sum_{k}\frac{1}{2C}(Q_{k}+q_{{\rm ind},k})^2.\label{HCdef}
\end{equation}
While both $E_{0}$ and $C$ may be different for different islands, we omit a possible $k$-dependence for ease of notation. There may be additional fluxes enclosed by the regions between the islands, but we do not include them to simplify the expressions. None of these simplifications is essential for the operation of the device.

The set of Majorana's on the $k$-th island is indicated by $\gamma_{kn}$ with $n=1,2,\ldots 2N_{k}$. The fermion parities ${\cal P}_{k}=i^{N_{k}}\prod_{n}\gamma_{kn}$ of neighboring islands $k$ and $k'$ are coupled with strength $E_{M}$ by the overlapping Majorana's $\gamma_{kn}$ and $\gamma_{k'm}$. We denote the gauge-invariant phase difference \cite{Tin96} by $\theta_{kk'}=\phi_{k}-\phi_{k'}+(2\pi/\Phi_{0})\int_{k\rightarrow k'}\bm{A}\cdot d\bm{l}$. The corresponding tunnel Hamiltonian \cite{Kit01}
\begin{equation}
H_{kk'}=\Gamma_{kk'}\cos(\theta_{kk'}/2),\;\;\Gamma_{kk'}=iE_{M}\gamma_{kn}\gamma_{k'm},\label{Hkkdef}
\end{equation}
is $4\pi$-periodic in the gauge-invariant phase difference, as an expression of the fact that single electrons (rather than Cooper pairs) tunnel through the midgap state. For example, in the two-island geometry of Fig.\ \ref{fig_islands} one has
\begin{subequations}
\label{H12def}
\begin{align}
&H_{12}=iE_{M}\gamma_{12}\gamma_{21}\cos(\theta_{12}/2),\label{H12defa}\\
&\theta_{12}=\phi_{1}-\phi_{2}-\pi(\Phi_{1}+\Phi_{2})/\Phi_{0}.\label{H12defb}
\end{align}
\end{subequations}

In Appendix \ref{Cinteraction} we derive the effective low-energy Hamiltonian in the regime $E_{J}\gg E_{C},E_{M}$,
\begin{align}
&H_{\rm eff}={\rm const}-\sum_{k}U_{k}{\cal P}_{k}+\sum_{k,k'}\Gamma_{kk'}\cos\alpha_{kk'},\label{Heffmulti}\\
&\alpha_{kk'}=\lim_{\phi_{k},\phi_{k'}\rightarrow 0}\tfrac{1}{2}\theta_{kk'}.\label{bardeltadef}
\end{align}
The single sum couples Majorana's within an island, through an effective Coulomb energy $U_{k}$. The double sum couples Majorana's in neighboring islands by tunneling. Both the Coulomb and tunnel couplings depend on the fluxes through the Josephson junctions, but in an entirely different way: the tunnel coupling varies slowly $\propto\cos(\pi\Phi/\Phi_{0})$ with the flux, while the Coulomb coupling varies rapidly $\propto\exp[-4\sqrt{(E_{0}/E_{C})\cos(\pi\Phi/\Phi_{0})}]$.

\subsection{Tri-junction}
\label{trijunction}

\begin{figure}[tb] 
\centerline{\includegraphics[width=0.8\linewidth]{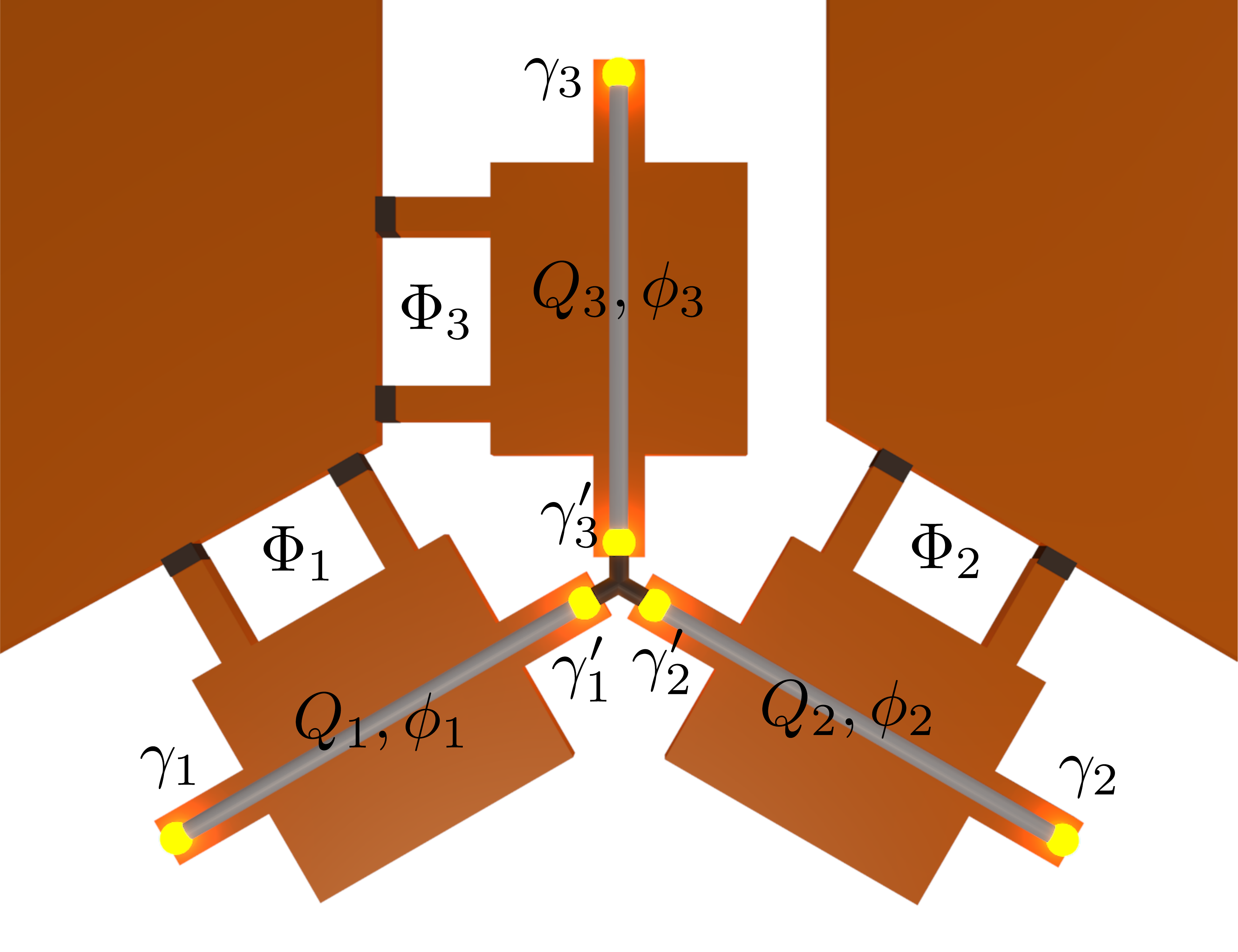}}
\caption{\label{fig_trijunction} 
Three Cooper pair boxes connected at a tri-junction via three overlapping Majorana fermions (which effectively produce a single zero-mode $\gamma_{0}$ at the center). This is the minimal setup required for the braiding of a pair of Majorana's, controlled by the fluxes through the three Josephson junctions to a bulk superconductor.
}
\end{figure}

Since ${\cal P}_{k}$ and $\Gamma_{kk'}$ in the Majorana-Coulomb Hamiltonian \eqref{Heffmulti} do not commute, the evolution of the eigenstates upon variation of the fluxes is nontrivial. As we will demonstrate, it can provide the non-Abelian braiding statistic that we are seeking. 

Similarly to earlier braiding proposals \cite{Ali11,Sau11}, the minimal setup consists of three superconductors in a tri-junction. (See Fig.\ \ref{fig_trijunction}.) Each superconductor contains a pair of Majorana fermions $\gamma_{k},\gamma'_{k}$, with a tunnel coupling between $\gamma'_{1},\gamma'_{2}$, and $\gamma'_{3}$. The Majorana-Coulomb Hamiltonian \eqref{Heffmulti} takes the form
\begin{align}
H_{\rm eff}={}&iE_M\bigl(\gamma'_1\gamma'_2\cos\alpha_{12}+\gamma'_2\gamma'_3\cos\alpha_{23}+\gamma'_3\gamma'_1\cos\alpha_{31}\bigr)\nonumber\\
&-\sum_{k=1}^3 U_k i\gamma_k\gamma'_k,\label{Hefftrijunction}
\end{align}
with gauge-invariant phase differences
\begin{subequations}
\label{alphatrijunction}
\begin{align}
&\alpha_{12}=-(\pi/2\Phi_{0})(\Phi_1+\Phi_2+2\Phi_{3}),\label{alphatrijunctiona}\\
&\alpha_{23}=(\pi/2\Phi_{0})(\Phi_2+\Phi_3),\label{alphatrijunctionb}\\
&\alpha_{31}=(\pi/2\Phi_{0})(\Phi_1+\Phi_3).\label{alphatrijunctionc}
\end{align}
\end{subequations}

As we vary $|\Phi_{k}|$ between $0$ and $\Phi_{\rm max}<\Phi_{0}/2$, the Coulomb coupling $U_{k}$ varies between two (possibly $k$-dependent) values $U_{\rm min}$ and $U_{\rm max}$. We require $U_{\rm max}\gg U_{\rm min}$, which is readily achievable because of the exponential flux sensitivity of the Coulomb coupling expressed by Eqs.\ \eqref{EJE0} and \eqref{Udef}. We call the Coulomb couplings $U_{\rm max}$ and $U_{\rm min}$ \textit{on} and \textit{off}, respectively. We also take $U_{\rm max}\ll E_{M}$, meaning that the Coulomb coupling is weaker than the tunnel coupling. This is not an essential assumption, but it allows us to reduce the 6--Majorana problem to a 4--Majorana problem, as we will now show.

Consider first the case that $U_k=0$ for all $k$. Then the Hamiltonian \eqref{Hefftrijunction} has four eigenvalues equal to zero: three of these represent the Majorana's $\gamma_{k}$ far away from the junction, while the fourth Majorana,
\begin{equation}
\gamma_0=\tfrac{1}{\sqrt{3}} (\gamma'_{1}+\gamma'_{2}+\gamma'_{3})\label{gamma0def}
\end{equation}
is situated at the tri-junction. The tri-junction contributes also two nonzero eigenvalues $\pm \frac{1}{2}E_{\rm gap}$, separated by the gap
\begin{equation}
E_{\rm gap}=E_{M}\sqrt{\cos^2\alpha_{12}+\cos^2\alpha_{23}+\cos^2\alpha_{31}}.\label{Egapdef}
\end{equation}
For $\Phi_{\max}$ well below $\Phi_{0}$ and $U_{\rm max}\ll E_{M}$ these two gapped modes can be ignored, and only the four Majorana's $\gamma_{0},\gamma_{1},\gamma_{2},\gamma_{3}$ need to be retained.

The Hamiltonian $H_{\rm int}$ that describes the Coulomb interaction of these four Majorana's for nonzero $U_{k}$ is given, to first order in $U_k/E_M$, by
\begin{align}
&H_{\rm int}=\sum_{k=1}^{3} \Delta_{k}\, i\gamma_0\gamma_k,\;\;\Delta_{k}=-(2E_{M}/E_{\rm gap})\beta_{k}U_{k},\label{Hint}\\
&\beta_{1}=\cos\alpha_{23},\;\;
\beta_{2}=\cos\alpha_{31},\;\;
\beta_{3}=\cos\alpha_{12}.
\label{betakdef}
\end{align}

\section{Majorana braiding}
\label{braiding}

The Hamiltonian \eqref{Hint} describes four flux-tunable Coulomb-coupled Majorana fermions. Although the coupling studied by Sau, Clarke, and Tewari \cite{Sau11} has an entirely different origin (gate-tunable tunnel coupling), their Hamiltonian has the same form. We can therefore directly adapt their braiding protocol to our control parameters.

\begin{figure}[tb] 
\centerline{\includegraphics[width=1\linewidth]{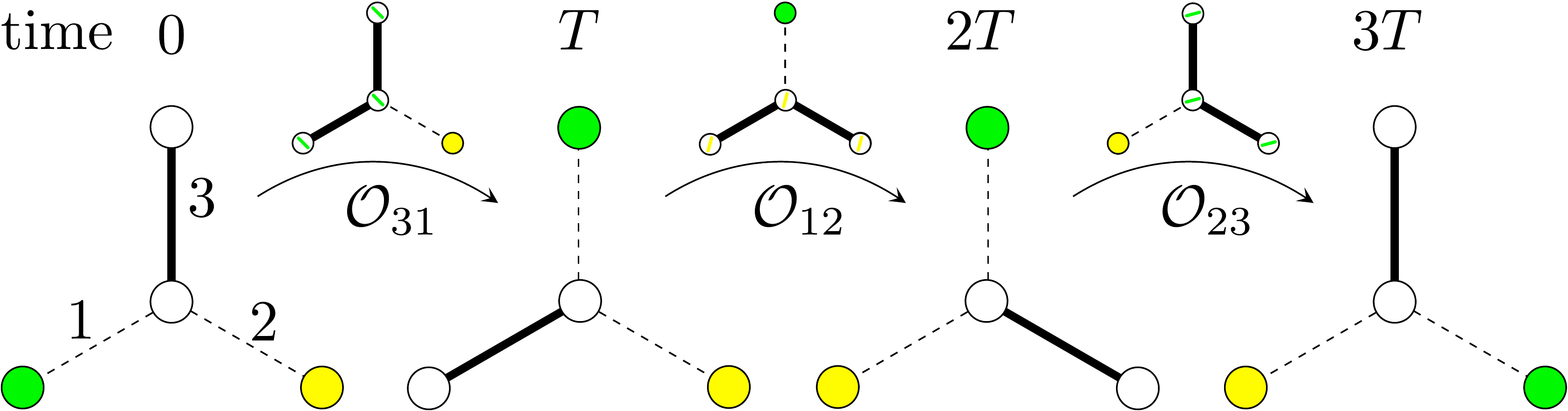}}
\caption{\label{fig_braiding} 
Schematic of the three steps of the braiding operation. The four Majorana's of the tri-junction in Fig.\ \ref{fig_trijunction} (the three outer Majorana's $\gamma_{1},\gamma_{2},\gamma_{3}$ and the effective central Majorana $\gamma_{0}$) are represented by circles and the Coulomb coupling is represented by lines (solid in the \textit{on} state, dashed in the \textit{off} state). White circles indicate Majorana's with a large Coulomb splitting, colored circles those with a vanishingly small Coulomb splitting. The small diagram above each arrow shows an intermediate stage, with one Majorana delocalized over three coupled sites. The three steps together exchange the Majorana's 1 and 2, which is a non-Abelian braiding operation.
}
\end{figure}

\begin{table}[tb]
\begin{center}
\begin{tabular}{l||ccc}
time & $\Phi_{1}$ & $\Phi_{2}$ & $\Phi_{3}$ \\
\hline\hline
$0$ & $0$ & $0$ & $-\Phi_{\rm max}$ \\
\hline
 & $\Phi_{\rm max}$ & $0$ & $-\Phi_{\rm max}$ \\
\hline
$T$ & $\Phi_{\rm max}$ & $0$ & $0$ \\
\hline
 & $\Phi_{\rm max}$& $\Phi_{\rm max}$ & $0$ \\
\hline
$2T$ & $0$ & $\Phi_{\rm max}$ & $0$ \\
\hline
 & $0$ & $\Phi_{\rm max}$ & $-\Phi_{\rm max}$ \\
\hline
$3T$ & $0$ & $0$ & $-\Phi_{\rm max}$ \\
\hline
\end{tabular}
\caption{\label{table1}
Variation of the flux through the three Josephson junctions during the braiding operation, at time steps corresponding to the diagrams in Fig.\ \ref{fig_braiding}. The flux $\Phi_{3}$ is varied in the opposite direction as $\Phi_{1},\Phi_{2}$, to ensure that the coupling parameters $\Delta_{k}\propto\beta_{k}$ do not change sign during the operation.
}
\end{center}
\end{table}

We have three fluxes $\Phi_{1},\Phi_{2},\Phi_{3}$ to control the couplings. The braiding operation consists of three steps, see Table \ref{table1} and Fig.\ \ref{fig_braiding}. (Ref.\ \cite{Sau11} had more steps, involving 6 rather than 4 Majorana's.) At the beginning and at the end of each step two of the couplings are \textit{off} ($\Phi_{k}=0$) and one coupling is \textit{on} ($|\Phi_{k}|=\Phi_{\rm max}$). We denote by ${\cal O}_{kk'}$ the step of the operation that switches the coupling that is \textit{on} from $k$ to $k'$. This is done by first increasing $|\Phi_{k'}|$ from $0$ to $\Phi_{\rm max}$ and then decreasing $|\Phi_{k}|$ from $\Phi_{\rm max}$ to $0$, keeping the third flux fixed at $0$. 

During this entire process the degeneracy of the ground state remains unchanged (twofold degenerate), which is a necessary condition for an adiabatic operation. If, instead, we would first have first decreased $|\Phi_{k}|$ and then increased $|\Phi_{k'}|$, the ground state degeneracy would have switched from two to four at some point during the process, precluding adiabaticity.

We start from coupling 3 \textit{on} and couplings 1,2 \textit{off}. The braiding operation then consists, in sequence, of the three steps ${\cal O}_{31}$, ${\cal O}_{12}$, and ${\cal O}_{23}$. Note that each coupling $\Delta_{k}$ appears twice in the \textit{on} state during the entire operation, both times with the same sign $s_{k}$.

The step ${\cal O}_{kk'}$ transfers the uncoupled Majorana at site $k'$ to site $k$ in a time $T$. The transfer is described in the Heisenberg representation by $\gamma_{k}(T)={\cal U}^{\dagger}(T)\gamma_{k}{\cal U}(T)$. We calculate the unitary evolution operator ${\cal U}(T)$ in the adiabatic $T\rightarrow\infty$ limit in Appendix \ref{Berry}, by integrating over the Berry connection. In the limit $U_{\rm min}\rightarrow 0$ we recover the result of Ref.\ \cite{Sau11},
\begin{equation}
\gamma_{k}(T)=-s_{k}s_{k'}\gamma_{k'}(0).\label{transfer_eq}
\end{equation}

The result after the three steps is that the Majorana's at sites 1 and 2 are switched, with a difference in sign,
\begin{equation}
\gamma_{1}(3T)=-s_{1}s_{2}\gamma_{2}(0),\;\;\gamma_{2}(3T)=s_{1}s_{2}\gamma_{1}(0).\label{gammaswitch}
\end{equation}
The corresponding unitary time evolution operator,
\begin{equation}
{\cal U}(3T)=\frac{1}{\sqrt 2}\bigl(1+s_{1}s_{2}\gamma_{1}\gamma_{2})=\exp\left(\frac{\pi}{4} s_{1}s_{2}\gamma_{1}\gamma_{2}\right),\label{U3T}
\end{equation}
has the usual form of an adiabatic braiding operation \cite{Iva01}. For a nonzero $U_{\rm min}$ the coefficient $\pi/4$ in the exponent acquires corrections of order $U_{\rm min}/U_{\rm max}$, see Appendix \ref{Berry}.

If one repeats the entire braiding operation, the Majorana's 1 and 2 have returned to their original positions but the final state differs from the initial state by a unitary operator ${\cal U}(3T)^{2}=s_{1}s_{2}\gamma_{1}\gamma_{2}$ and not just by a phase factor. That is the hallmark of non-Abelian statistics \cite{Rea00}.

\section{Discussion}
\label{discuss}

In summary, we have proposed a way to perform non-Abelian braiding operations on Majorana fermions, by controlling their Coulomb coupling via the magnetic flux through a Josephson junction. Majorana fermions are themselves charge-neutral particles (because they are their own antiparticle), so one may ask how there can be any Coulomb coupling at all. The answer is that the state of a pair of Majorana fermions in a superconducting island depends on the parity of the number of electrons on that island, and it is this dependence on the electrical charge modulo $2e$ which provides an electromagnetic handle on the Majorana's.

The Coulomb coupling can be made exponentially small by passing Cooper pairs through a Josephson junction between the island and a bulk (grounded) superconductor. The control parameter is the flux $\Phi$ through the junction, so it is purely magnetic. This is a key difference with braiding by electrostatically controlled tunnel couplings of Majorana fermions \cite{Sau11}. Gate voltages tend to be screened quite efficiently by the superconductor, so magnetic control is advantageous. Another advantage is that the dependence of the Coulomb coupling on the flux is governed by macroscopic electrical properties (capacitance of the island, resistance of the Josephson junction). Tunnel couplings, in contrast, require microscopic input (separation of the Majorana fermions on the scale of the Fermi wave length), so they tend to be more difficult to control.

Both Ref.\ \cite{Sau11} and the present proposal share the feature that the gap of the topological superconductor is not closed during the braiding operation. (The measurement-based approach to braiding also falls in this category \cite{Bon08}.) Two other proposals \cite{Ali11,Rom11} braid the Majorana's by inducing a topological phase transition (either by electrical or by magnetic means) in parts of the system. Since the excitation gap closes at the phase transition, this may be problematic for the required adiabaticity of the operation. 

The braiding operation is called topologically protected, because it depends on the \textit{off/on} sequence of the Coulomb couplings, and not on details of the timing of the sequence. As in any physical realization of a mathematical concept, there are sources of error. Non-adiabaticity of the operation is one source of error, studied in Ref.\ \cite{Che11}. Low-lying sub-gap excitations in the superconducting island break the adiabatic evolution by transitions which change the fermion parity of the Majorana's.

Another source of error, studied in Appendix \ref{Berry}, is governed by the \textit{off/on} ratio $U_{\rm min}/U_{\rm max}$ of the Coulomb coupling. This ratio depends exponentially on the ratio of the charging energy $E_C$ and the Josephson energy $E_J$ of the junction to the bulk superconductor. A value $E_{J}/E_{C}\simeq 50$ is not unrealistic \cite{Sch08}, corresponding to $U_{\rm min}/U_{\rm max}\simeq 10^{-5}$.

The sign of the Coulomb coupling in the \textit{on} state can be arbitrary, as long as it does not change during the braiding operation. Since $U_{\rm max}\propto\cos(\pi q_{\rm ind}/e)$, any change in the induced charge by $\pm e$ will spoil the operation. The time scale for this quasiparticle poisoning can be milliseconds \cite{Vis11}, so this does not seem to present a serious obstacle.

A universal quantum computation using Majorana fermions requires, in addition to braiding, the capabilities for single-qubit rotation and read-out of up to four Majorana's \cite{Nay08}. The combination of Ref.\ \cite{Has11} with the present proposal provides a scheme for all three operations, based on the interface of a topological qubit and a superconducting charge qubit. This is not a topological quantum computer, since single-qubit rotations of Majorana fermions lack topological protection. But by including the topologically protected braiding operations one can improve the tolerance for errors of the entire computation by orders of magnitude (error rates as large as 10\% are permitted \cite{Bra05}).

\begin{figure}[tb] 
\centerline{\includegraphics[width=1\linewidth]{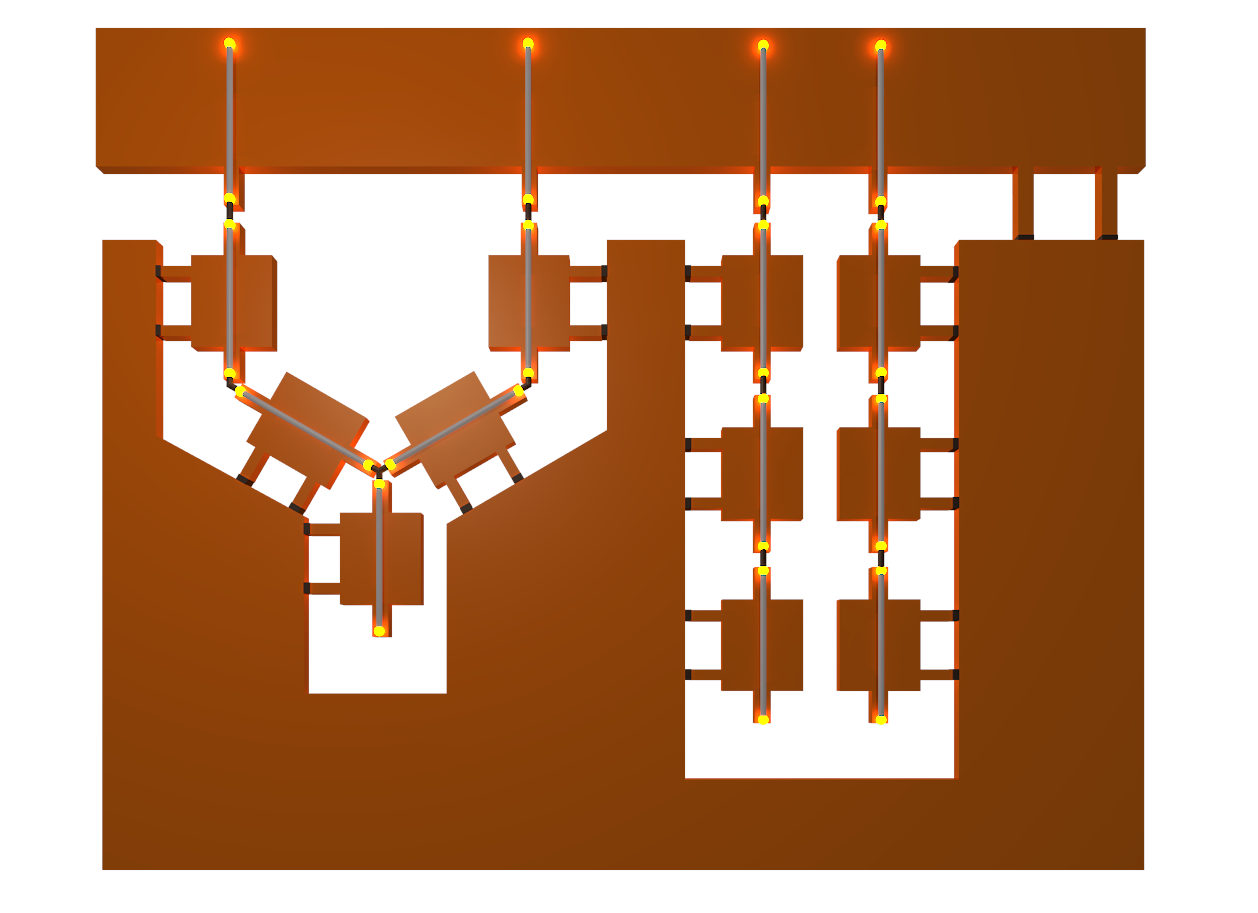}}
\caption{\label{fig_device} 
Josephson junction array containing Majorana fermions. The magnetic flux through a split Josephson junction controls the Coulomb coupling on each superconducting island. This device allows one to perform the three types of operations on topological qubits needed for a universal quantum computer: read-out, rotation, and braiding. All operations are controlled magnetically, no gate voltages are needed.
}
\end{figure}

A sketch of a complete device is shown in Fig.\ \ref{fig_device}.

\acknowledgments

This research was supported by the Dutch Science Foundation NWO/FOM and by an ERC Advanced Investigator Grant.

\appendix

\section{Derivation of the Majorana-Coulomb Hamiltonian}
\label{Cinteraction}

\subsection{Single island}
\label{Csingle}

Considering first a single island, we start from the Cooper pair box Hamiltonian \eqref{Hsingle} with the parity constraint \eqref{Psiphi} on the eigenstates. Following Ref.\ \cite{Hec11}, it is convenient to remove the constraint by the unitary transformation
\begin{equation}
\tilde{H}=\Omega^\dagger H \Omega,\;\;\Omega=\exp[i(1-{\cal P}) \phi/4].\label{gaugetr}
\end{equation}
The transformed wave function $\tilde{\Psi}(\phi)=\Omega^{\dagger}\Psi(\phi)$ is then $2\pi$-periodic, without any constraint. The parity operator ${\cal P}$ appears in the transformed Hamiltonian,
\begin{equation}
\tilde{H}=\frac{1}{2C}\bigl(Q+\tfrac{1}{2}e(1-{\cal P})+q_{\rm ind}\bigr)^2-E_J\cos\phi.\label{tildeH}
\end{equation}

For a single junction the parity is conserved, so eigenstates of $H$ are also eigenstates of ${\cal P}$ and we may treat the operator ${\cal P}$ as a number. Eq.\ \eqref{tildeH} is therefore the  Hamiltonian of a Cooper pair box with effective induced charge $q_{\rm eff}=q_{\rm ind}+e(1-{\cal P})/2$. The expression for the ground state energy in the Josephson regime $E_{J}\gg E_{C}$ is in the literature \cite{Koc07,Naz09},
\begin{align}
&E_{\rm ground}=-E_{J}+\sqrt{2 E_CE_J}\nonumber\\
&\quad-16(E_{C}E_{J}^{3}/2\pi^{2})^{1/4}\,e^{-\sqrt{8E_{J}/E_{C}}}\,\cos(\pi q_{\rm eff}/e).\label{WKBshift}
\end{align}

The first term $-E_J$ is the minimal Josephson energy at $\phi_{\rm min}=0$. Zero-point motion, with Josephson plasma frequency $\omega_{p}=\sqrt{8E_{C}E_{J}}/\hbar$, adds the second term $\sqrt{2E_{C}C_{J}}=\frac{1}{2}\hbar\omega_{p}$. The third term is due to quantum phase slips with transition amplitudes $\tau_{\pm}\simeq \exp(\pm i \pi q_{\rm eff}/e)\sqrt{\hbar\omega_{p}E_{J}}\exp(-\hbar\omega_{p}/E_{J})$ by which $\phi$ increments by $\pm 2\pi$. 

Using ${\cal P}^{2}=1$, the ground state energy \eqref{WKBshift} may be written in the form
\begin{equation}
E_{\rm ground}=-E_J+\sqrt{2 E_C E_J} -U {\cal P},\label{Eground}
\end{equation}
with $U$ defined in Eq.\ \eqref{Udef}. Higher levels are separated by an energy $\hbar\omega_{p}$, which is large compared to $U$ for $E_{J}\gg E_{C}$. We may therefore identify $E_{\rm ground}=H_{\rm eff}$ with the effective low-energy Hamiltonian of a single island in the large-$E_{J}$ limit.

\subsection{Multiple islands}
\label{Cmultiple}

We now turn to the case of multiple islands with tunnel coupling. To be definite we take the geometry of two islands shown in Fig.\ \ref{fig_islands}. The full Hamiltonian is $H=H_1+H_2+H_{12}$, where $H_1$ and $H_2$ are two copies of the Cooper box Hamiltonian \eqref{Hsingle} and $H_{12}$ is the tunnel coupling from Eq.\ \eqref{H12def}. 

To obtain $2\pi$-periodicity in both phases $\phi_{1}$ and $\phi_{2}$, we make the unitary transformation $\tilde{H}=\Omega^\dagger H \Omega$ with
\begin{equation}
\Omega=e^{i(1-{\cal P}_1) \phi_1/4}e^{i(1-{\cal P}_2) \phi_2/4}.\label{Omega12def}
\end{equation}
The Cooper pair box Hamiltonians are transformed into
\begin{equation}
\tilde{H}_{k}=\frac{1}{2C}\bigl(Q_{k}+eq_{k}+q_{{\rm ind},k}\bigr)^2-E_{J,k}\cos\phi_{k},\label{tildeHk}
\end{equation}
with $q_{k}=\tfrac{1}{2}(1-{\cal P}_{k})$. The tunnel coupling transforms into
\begin{equation}
\tilde{H}_{12}=\tfrac{1}{2}e^{-iq_1\phi_1}\Gamma_{12}e^{iq_2\phi_2}\,e^{i\pi(\Phi_1+\Phi_2)/2\Phi_0}+\text{H.c.},\label{tildeH12}
\end{equation}
where $\Gamma_{12}=iE_{M}\gamma_{12}\gamma_{21}$ and H.c.\ stands for Hermitian conjugate. Since $e^{iq\phi}=\cos\phi+iq\sin\phi$, the transformed tunnel coupling $\tilde{H}_{12}$ is $2\pi$-periodic in $\phi_{1}$ and $\phi_{2}$.

For $E_{J}\gg E_{C}$ the phases remain close to the value which minimizes the sum of the Josephson energies to the bulk superconductor and between the islands. To leading order in $E_{M}/E_{J}\ll 1$ this minimal energy is given by
\begin{align}
{\cal E}_{\rm min}={}&-E_{J,1}-E_{J,2}+\Gamma_{12}\cos[\pi(\Phi_{1}+\Phi_{2})/2\Phi_{0}]\nonumber\\
&+{\cal O}(E_{M}^{2}/E_{J}).\label{Emindef}
\end{align}
The Josephson coupling of the islands changes the plasma frequency $\omega_{p,k}$ for phase $\phi_{k}$ by a factor $1+{\cal O}(E_{M}/E_{J})$, so the zero-point motion energy is
\begin{equation}
\tfrac{1}{2}\hbar\omega_{p,k}=\sqrt{2 E_{C}E_{J,k}}+E_{M}\times{\cal O}(E_{C}/E_{J})^{1/2}.\label{omegapcorrected}
\end{equation}
The transition amplitudes $\tau_{\pm}$ for quantum phase slips of phase $\phi_{k}$ are similarly affected,
\begin{equation}
\tau_{\pm,k}=-U_{k}{\cal P}_{k}+E_{M}e^{-\hbar\omega_{p,k}/E_{J,k}}\times{\cal O}(E_{C}/E_{J})^{1/4}.\label{taupmcorrected}
\end{equation}

These are the contributions to the effective Hamiltonian $H_{\rm eff}={\cal E}_{\rm min}+\sum_{k}(\frac{1}{2}\hbar\omega_{p,k}+\tau_{+,k}+\tau_{-,k})$ for $E_{J}\gg E_{C},E_{M}$,
\begin{align}
H_{\rm eff}={}&\biggl(-U_{1}{\cal P}_{1}-U_{2}{\cal P}_{2}+\Gamma_{12}\cos[\pi(\Phi_{1}+\Phi_{2})/2\Phi_{0}]\biggr)\nonumber\\
&\times[1+{\cal O}(E_{M}/E_{J})]+{\rm const}.\label{Heffmultiapp}
\end{align}
Eq.\ \eqref{Heffmulti} in the main text generalizes this expression for two islands to an arbitrary number of coupled islands.

\section{Calculation of the Berry phase of the braiding operation}
\label{Berry}

We evaluate the unitary evolution operator ${\cal U}$ of the braiding operation in the adiabatic limit. This amounts to a calculation of the non-Abelian Berry phase (integral of Berry connection) of the cyclic variation of the interaction Hamiltonian $H_{\rm int}(\Delta_{1},\Delta_{2},\Delta_{3})$.

In the Fock basis $|00\rangle, |01\rangle, |10\rangle, |11\rangle$ the interaction Hamiltonian \eqref{Hint} of 4 Majorana fermions is given by the occupation number of the two fermionic operators $c_1=\left( \gamma_1-i\gamma_2\right)/2$ and $c_2=\left(\gamma_0-i\gamma_3 \right)/2$. It takes the form
\begin{widetext}
\begin{equation}
H_{\rm int}=\left(\begin{array}{cccc} -\Delta_3 & 0 & 0 & -i\Delta_1-\Delta_2\\ 0 & \Delta_3 & -i\Delta_1-\Delta_2 & 0 \\ 0& i\Delta_1-\Delta_2 & -\Delta_3 & 0 \\ i\Delta_1-\Delta_2 & 0 & 0 & \Delta_3\end{array}\right).\label{HintFock}
\end{equation}
\end{widetext}
The eigenvalues are doubly degenerate at energy $\pm\varepsilon=\pm\sqrt{\Delta_1^2+\Delta_2^2+\Delta_3^2}$ (up to a flux-dependent offset, which only contributes an overall phase factor to the evolution operator). The two degenerate ground states at $-\varepsilon$ are distinguished by an even ($e$) or odd ($o$) quasiparticle number,
\begin{subequations}
\label{param}
\begin{align}
&|e\rangle=\sqrt{\frac{\varepsilon-\Delta_3}{2\varepsilon}}\left(\begin{array}{c} i\dfrac{\varepsilon+\Delta_3}{\Delta_1+i\Delta_2} \\0 \\0 \\1\end{array}\right),\\
&|o\rangle=\sqrt{\frac{\varepsilon+\Delta_3}{2\varepsilon}}\left(\begin{array}{c} 0\\ i\dfrac{\varepsilon-\Delta_3}{\Delta_1+i\Delta_2}  \\1 \\0\end{array}\right).
\end{align}
\end{subequations}
This parameterization is smooth and continuous except along the line $\Delta_1=\Delta_2=0$. 

\begin{figure}[tb] 
\centerline{\includegraphics[width=0.7\linewidth]{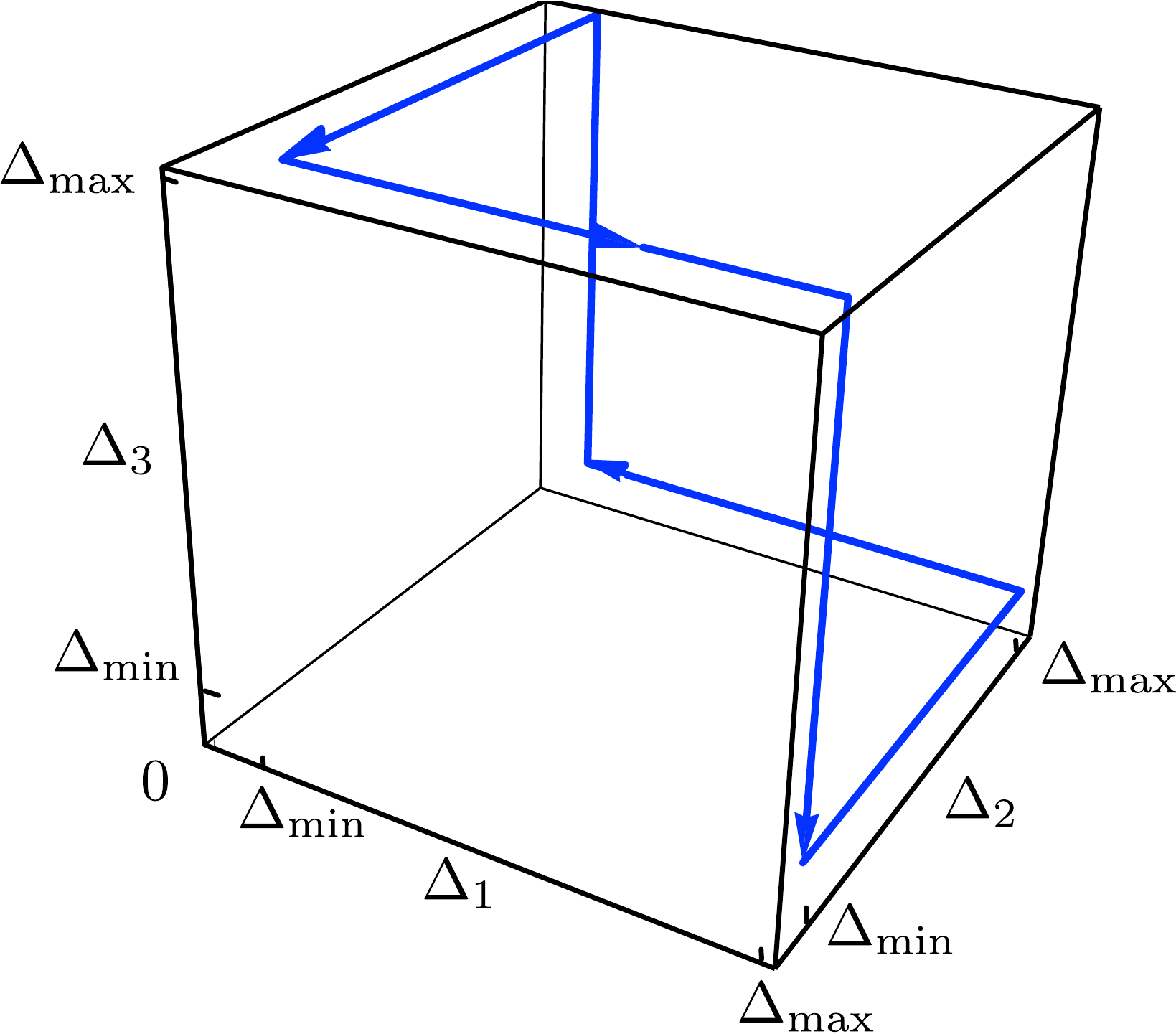}}
\caption{\label{fig_path} 
The braiding path in three-dimensional parameter space along which the Berry phase is evaluated. This path corresponds to the flux values in Table \ref{table1}, with couplings $\Delta_{k}=\Delta_{\rm min}$ for $\Phi_{k}=0$ and $\Delta_{k}=\Delta_{\rm max}$ for $|\Phi_{k}|=\Phi_{\rm max}$. The ratio $\Delta_{\rm min}/\Delta_{\rm max}$ in the figure is exaggerated for clarity.
}
\end{figure}

If we avoid this line the Berry connection can be readily evaluated. It consists of three anti-Hermitian $2\times 2$ matrices ${\cal A}_k$,
\begin{equation}
{\cal A}_k=\left(\begin{array}{cc} \langle e |\,\frac{d}{d \Delta_k}\,| e \rangle & 0 \\ 0 & \langle o |\,\frac{d}{d \Delta_k}\,| o \rangle \end{array}\right).
\end{equation}
Off-diagonal terms in ${\cal A}_k$ are zero because of global parity conservation. Explicitly, we have
\begin{align}
&{\cal A}_1=\frac{\Delta_2}{\Delta_1^2+\Delta_2^2}\,\left(\begin{array}{cc} i \,\dfrac{\varepsilon+\Delta_3}{2\varepsilon} & 0 \\ 0 & i\,\dfrac{\varepsilon-\Delta_3}{2\varepsilon}  \end{array}\right),\\
&{\cal A}_2=\frac{-\Delta_1}{\Delta_1^2+\Delta_2^2}\,\left(\begin{array}{cc} i \,\dfrac{\varepsilon+\Delta_3}{2\varepsilon} & 0 \\ 0 & i\,\dfrac{\varepsilon-\Delta_3}{2\varepsilon}  \end{array}\right),\\
&{\cal A}_3=0.
\end{align}

A closed path ${\cal C}$ in parameter space has Berry phase \cite{Wil84}
\begin{equation}
{\cal U}=\exp\left(-\oint_{{\cal C}}\sum_{k}A_k \,d\Delta_k\right).\label{Berryphase}
\end{equation}
The path ${\cal C}$ corresponding to the braiding operation in Fig.\ \ref{fig_braiding} and Table \ref{table1} is shown in Fig.\ \ref{fig_path}. We take all couplings $\Delta_{k}$ positive, varying between a minimal value $\Delta_{\rm min}$ and maximal value $\Delta_{\rm max}$. The parametrization \eqref{param} is well-defined along the entire contour. 

The contour integral evaluates to 
\begin{align}
&{\cal U}=\exp\left[-i\left(\frac{\pi}{4}-\epsilon\right) \sigma_z\right],\;\;\sigma_{z}=\begin{pmatrix}
1&0\\
0&-1
\end{pmatrix},\\
&\epsilon=\frac{3}{\sqrt{2}}\frac{\Delta_{\rm min}}{\Delta_{\rm max}}+{\cal O}\left(\frac{\Delta_{\rm min}}{\Delta_{\rm max}}\right)^{2}. \label{braiding_error}
\end{align}
The limit $\Delta_{\rm min}/\Delta_{\rm max}\rightarrow 0$ corresponds to the braiding operator \eqref{U3T} in the main text (with $s_{1},s_{2}>0$ and $\sigma_{z}=1-2c_{1}^{\dagger}c_{1}=i\gamma_{1}\gamma_{2}$).


\begin{thebibliography}{99}
\bibitem{Nay08} C. Nayak, S. Simon, A. Stern, M. Freedman, and S. Das Sarma, Rev. Mod. Phys. \textbf{80}, 1083 (2008).
\bibitem{Lut10} R. M. Lutchyn, J. D. Sau, and S. Das Sarma, Phys. Rev. Lett. \textbf{105}, 077001 (2010).
\bibitem{Ore10} Y. Oreg, G. Refael, and F. von Oppen, Phys. Rev. Lett. \textbf{105}, 177002 (2010).
\bibitem{Kit01} A. Yu. Kitaev, Phys. Usp. \textbf{44} (suppl.), 131 (2001).
\bibitem{Ave92} D. V. Averin and Yu. V. Nazarov, Phys. Rev. Lett. \textbf{69}, 1993 (1992).
\bibitem{Dev04} M. H. Devoret, A. Wallraff, and J. M. Martinis, arXiv:cond-mat/0411174.
\bibitem{Sch08} J. A. Schreier, A. A. Houck, J. Koch, D. I. Schuster, B. R. Johnson, J. M. Chow, J. M. Gambetta, J. Majer, L. Frunzio, M. H. Devoret, S. M. Girvin, and R. J. Schoelkop, Phys. Rev. B \textbf{77}, 180502(R) (2008).
\bibitem{Has11} F. Hassler, A. R. Akhmerov, and C. W. J. Beenakker, New J. Phys. \textbf{13}, 095004 (2011).
\bibitem{Bra05} S. Bravyi and A. Yu. Kitaev, Phys. Rev. A \textbf{71}, 022316 (2005); S. Bravyi, Phys. Rev. A \textbf{73}. 042313 (2006).
\bibitem{Rea00} N. Read and D. Green, Phys. Rev. B \textbf{61}, 10267 (2000).
\bibitem{Iva01} D. Ivanov, Phys. Rev. Lett. \textbf{86}, 268 (2001).
\bibitem{Has10} F. Hassler, A. R. Akhmerov, C.-Y. Hou, and C. W. J. Beenakker, New J. Phys. \textbf{12}, 125002 (2010).
\bibitem{Sau10} J. D. Sau, S. Tewari, and S. Das Sarma, Phys. Rev. A \textbf{82}, 052322 (2010).
\bibitem{Fle11} K. Flensberg, Phys. Rev. Lett. \textbf{106}, 090503 (2011).
\bibitem{Jia11} L. Jiang, C. L. Kane, and J. Preskill, Phys. Rev. Lett. \textbf{106}, 130504 (2011).
\bibitem{Bon11} P. Bonderson and R. M. Lutchyn, Phys. Rev. Lett. \textbf{106}, 130505 (2011).
\bibitem{Ali11} J. Alicea, Y. Oreg, G. Refael, F. von Oppen, and M. P. A. Fisher, Nature Phys. \textbf{7}, 412 (2011).
\bibitem{Sau11} J. D. Sau, D. J. Clarke, and S. Tewari, Phys. Rev. B \textbf{84}, 094505 (2011).
\bibitem{Rom11} A. Romito, J. Alicea, G. Refael, and F. von Oppen, arXiv:1110.6193.
\bibitem{Fu10} L. Fu, Phys. Rev. Lett. \textbf{104}, 056402 (2010); C. Xu and L. Fu, Phys. Rev. B \textbf{81}, 134435 (2010).
\bibitem{Hec11} B. van Heck, F. Hassler, A. R. Akhmerov, and C. W. J. Beenakker, Phys. Rev. B \textbf{4}, 180502(R) (2011).
\bibitem{Wil84} F. Wilczek and A. Zee, Phys. Rev. Lett. \textbf{52}, 2111 (1984).
\bibitem{Mak01} Yu. Makhlin, G. Sch\"{o}n, and A. Shnirman, Rev. Mod. Phys. \textbf{73}, 357 (2001).
\bibitem{Tin96} M. Tinkham, \textit{Introduction to Superconductivity} (McGraw-Hill, New York, 1996).
\bibitem{Bon08} P. Bonderson, M. Freedman, and C. Nayak, Phys. Rev. Lett. \textbf{101}, 010501 (2008).
\bibitem{Che11} M. Cheng, V. Galitski, and S. Das Sarma, Phys. Rev. B \textbf{84}, 104529 (2011).
\bibitem{Vis11} P. J. de Visser, J. J. A. Baselmans, P. Diener, S. J. C. Yates, A. Endo, and T. M. Klapwijk, Phys. Rev. Lett. \textbf{106}, 167004 (2011).
% references from Appendices
\bibitem{Koc07} J. Koch, T. M. Yu, J. Gambetta, A. A. Houck, D. I. Schuster, J. Majer, A. Blais, M. H. Devoret, S. M. Girvin, and R. J. Schoelkopf, Phys. Rev. A \textbf{76}, 042319 (2007).
\bibitem{Naz09} Yu. V. Nazarov and Ya. M. Blanter, \textit{Quantum Transport: Introduction to Nanoscience} (Cambridge, 2009).
\end{thebibliography}
\end{document}